# Computing Reformulated First Zagreb Index of Some Chemical Graphs as an Application of Generalized Hierarchical Product of Graphs

Nilanjan De
Calcutta Institute of Engineering and Management, Kolkata, India. E-mail: de.nilanjan@rediffmail.com

**ABSTRACT**

*The generalized hierarchical product of graphs was introduced by L. Barriére et al in 2009. In this paper, reformulated first Zagreb index of generalized hierarchical product of two connected graphs and hence as a special case cluster product of graphs are obtained. Finally using the derived results the reformulated first Zagreb index of some chemically important graphs such as square comb lattice, hexagonal chain, molecular graph of truncated cube, dimer fullerene, zig-zag polyhex nanotube and dicentric dendrimers are computed.*

## 1. INTRODUCTION

In Let $G$ be a simple connected graph with vertex set $V(G)$ and edge set $E(G)$. Let $n$ and $m$ respectively denote the order and size of $G$. In this paper we consider only simple connected graph, that is, graphs without any self loop or parallel edges. In molecular graph theory, molecular graphs represent the chemical structures of a chemical compound and it is often found that there is a correlation between the molecular structure descriptor with different physico-chemical properties of the corresponding chemical compounds. These molecular structure descriptors are commonly known as topological indices which are some numeric parameter obtained from the molecular graphs and are necessarily invariant under automorphism. Thus topological indices are very important useful tool to discriminate isomers and also shown its applicability in quantitative structure-activity relationship (QSAR), structure-property relationship (QSPR) and nanotechnology including discovery and design of new drugs.

Among different types of topological indices, the vertex degree based topological indices are very useful topological indices in study of structure property correlation of molecules. The degree of a vertex $v$ of a graph $G$ is equal to number of vertices adjacent with it and is denoted by $d(v)$ whereas $\delta(v)$ represents the sum of degrees of all the vertices adjacent to $v$. The first and second Zagreb indices are the oldest vertex degree based topological indices and was introduced by Gutman and Trinajstić [1] and are defined as

$$M_1(G) = \sum_{v \in V(G)} d(u)^2 = \sum_{uv \in E(G)} [d(u) + d(v)] \text{ and } M_2(G) = \sum_{uv \in E(G)} d(u)d(v).$$

These indices have been extensively studied both with respect to mathematical and chemical point of view. There are various extension of Zagreb indices. In [2], Milićević et al. Introduced a new version of Zagreb indices called reformulated Zagreb index. Here the degree of a vertex of classical Zagreb indices is replaced by degree of an edge e denoted by $d(e)$, where the degree of an edge $e = uv$ is given by $d(u) + d(v) - 2$. Therefore, the first and second reformulated Zagreb indices of a graph $G$ are defined as

$$EM_1(G) = \sum_{e \in E(G)} d(e)^2 = \sum_{e=uv \in E(G)} \{d(u) + d(v) - 2\}^2$$

$$EM_2(G) = \sum_{e \sim f \in E(G)} d(e)d(f).$$

In the definition of second reformulated Zagreb index, the notation $e \sim f$ means that the edges $e$ and $f$ are adjacent, that is, they share a common vertex in $G$. There are a large number of studies regarding various chemical and mathematical properties of the reformulated Zagreb indices. Different basic properties and bounds of reformulated Zagreb indices have been studied in [3] and [4]. In [5], bounds for the reformulated first Zagreb index of graphs with connectivity at most $k$ are obtained. De [6], found some upper and lower bounds of these indices in terms of some other graph invariants and also derived reformulated Zagreb indices of a class of dendrimers in [7]. Ji et al. in [8] and [9], computed these indices for acyclic, unicyclic, bicyclic and tricyclic graphs. Recently De et al. [10], investigate reformulated first Zagreb index of various graph operations.

We know that, different graph operations create a new graph from the given simpler graphs and sometimes it is found that, some chemically interesting graphs can be obtained as a result of graph operations of some simpler graphs. So from the relations obtained for different topological indices of graph operations in terms of topological indices of their components, it is convenient to find topological indices of some special molecular graphs and nanostructures. There are several study concerning topological indices of different graph operations. In [11], [12] and [13], De et al. obtained connective eccentric index, F-index and F-coindex of different graph operations respectively. Khalifeh et al., in [14] obtained some exact expressions for computing first and second Zagreb indices of different graph operations. In [15], Ashrafi et al. compute expressions for Zagreb coindices of different graph operations. In [16], Das et al. found some upper bounds for multiplicative Zagreb indices of some graph operations and Azari et al. in [17], derived explicit formulae for computing the eccentric-distance sum of different graph operations. Recently, De computes the vertex Zagreb indices of different graph operations in [18]. For more results, interested readers are referred to the papers [19-22]. In this study, our goal is to study reformulated first Zagreb index of an important graph operation called generalised hierarchical product of graphs to obtain this topological index of different chemical graphs. Barriére et al. in 2009 introduced the generalised hierarchical product of graphs [23],which is a generalization of both Cartesian product of graphs and the (standard) hierarchical product of graphs [24]. There are various studies of generalised hierarchical product of graphs for different topological indices in recent literature [25-28].

In this paper, we first derive explicit expression of reformulated first Zagreb index of generalized hierarchical product of two connected graphs. Hence using the derived results, the reformulated first Zagreb index of some chemically important graphs such as square comb lattice, hexagonal chain, molecular graph of truncated cube, dimer fullerene etc. are obtained.

## 2. GENERALISED HIERARCHICAL PRODUCT OF GRAPHS

The generalised hierarchical product of graphs is one of most important graph operations as many other graph operations such as Cartesian product of graphs, cluster product of graphs can be considered as a special case of this graph operation. There are already various studies on different topological indices of this graph operation till date.

**Definition.** Let $G$ and $H$ be two connected graphs and $U$ be a non-empty subset of $V(H)$. Then the hierarchical product of $G$ and $H$, denoted by $G \sqcap H(U)$, is the graph with vertex set $V(G) \times V(H)$, and any two vertices $(u,v)$ and $(u',v')$ of $G \sqcap H(U)$ are adjoint by an edge if and only if $\left[ v = v' \in U \text{ and } uu' \in E(G) \right]$ or $\left[ v = v' \text{ and } vv' \in E(H) \right]$.

From definition, we can state the following lemma which gives some basic properties of generalized hierarchical product of graphs.

**Lemma.1** Let $G$ and $H$ be graphs with $U \subseteq V(H)$. Then
  (a) $G\Pi H(U)$ is connected if and only if $G$ and $H$ are connected.
  (b) The degree of a vertex $(a,x)$ of $G\Pi H(U)$ is given by

$$d_{G(U)\Pi H}(a,x) = \begin{cases} d(x) + \chi_U(x)d(a), a \in V(G), x \in E(H) \\ d(x) + d(a), a \in V(G), x \in U \end{cases}$$

where, $\chi_U(x)$ is characteristic function on the set $U$, which is 1 on $U$ and 0 outside $U$.

Now let us first derive the reformulated first Zagreb index of generalised hierarchical product of graphs $G$ and $H$.

**Theorem.1** The reformulated first Zagreb index of $G\Pi H(U)$ is given by

$$EM_1(G\Pi H(U)) = |V(G)|EM_1(H) + |U|EM_1(G) + 5M_1(G)\sum_{u \in U} d(u) + 8|E(G)|\sum_{u \in U} d(u)^2$$

$$+ 2M_1(G)\{xy \in E(H) : x, y \in U\} + 4|E(G)|\sum_{u \in U}\sum_{x \in N[u]} d(x) - 16|E(G)|\sum_{u \in U} d(u).$$

**Proof.** Let $G$ and $H$ be two connected graphs and $U$ be a nonempty subset of $V(H)$. Let us partition the edge set of $G\Pi H(U)$ into two subsets, say $E_1$ and $E_2$ so that

$E_1 = \{(a,x)(b,y) : xy \in E(H) \text{ and } a = b \in V(G)\}$

$E_2 = \{(a,x)(b,y) : ab \in E(G) \text{ and } x = y \in V(H)\}$.

Now using definition of reformulated first Zagreb index and generalised hierarchical product graph, we consider the following two cases to calculate the reformulated first Zagreb index of $G\Pi H(U)$.

**Case. 1** The contributions of the edges of $E_1$ to the reformulated first Zagreb index of $G\Pi H(U)$ is calculated as

$$H_1 = \sum_{a \in V(G), xy \in E(H)} \{d(a,x) + d(a,y) - 2\}^2$$

$$= \sum_{a \in V(G), xy \in E(H)} \{d(x) + \chi_U(x)d(a) + d(y) + \chi_U(y)d(a) - 2\}^2$$

$$= \sum_{a \in V(G), xy \in E(H)} \{d(x) + d(y) - 2\}^2 + \sum_{a \in V(G), xy \in E(H)} \{\chi_U(x)d(a) + \chi_U(y)d(a)\}^2$$

$$+ 2\sum_{a \in V(G), xy \in E(H)} d(a)\{d(x) + d(y) - 2\}\{\chi_U(x) + \chi_U(y)\}$$

$$= |V(G)|EM_1(H) + \sum_{a \in V(G), xy \in E(H)} d(a)^2\{\chi_U(x)^2 + \chi_U(y)^2 + 2\chi_U(x)\chi_U(y)\}$$

$$+ 2\sum_{a \in V(G), xy \in E(H)} d(a)\{(\chi_U(x)d(x) + \chi_U(y)d(y)) + (\chi_U(x)d(y) + \chi_U(y)d(x)) - 2(\chi_U(x) + \chi_U(y))\}$$

$$= |V(G)|EM_1(H) + \sum_{a \in V(G), xy \in E(H)} d(a)^2\{\chi_U(x)^2 + \chi_U(y)^2\} + 2\sum_{a \in V(G), xy \in E(H)} d(a)^2 \chi_U(x)\chi_U(y)$$

$$+ 2\sum_{a \in V(G), xy \in E(H)} d(a)\{\chi_U(x)d(x) + \chi_U(y)d(y)\} + 2\sum_{a \in V(G), xy \in E(H)} d(a)\{\chi_U(x)d(y) + \chi_U(y)d(x)\}$$

$$- 4\sum_{a \in V(G), xy \in E(H)} d(a)\{\chi_U(x) + \chi_U(y)\}$$

$$= |V(G)| EM_1(H) + M_1(G) \sum_{u \in U} d(u) + 2M_1(G) |\{xy \in E(G) : x, y \in U\}|$$
$$+ 4|E(G)| \sum_{u \in U} d(u)^2 + 4|E(G)| \sum_{u \in U} \sum_{x \in N[u]} d(x) - 8|E(G)| \sum_{u \in U} d(u).$$

Case. 2 Similarly, the contributions of the edges of $E_2$ to the reformulated first Zagreb index of $G \Pi H(U)$ is calculated as

$$H_2 = \sum_{ab \in E(G), x \in U} \{d(a, x) + d(b, x) - 2\}^2$$

$$= \sum_{ab \in E(G), x \in U} \{d(x) + d(a) + d(x) + d(b) - 2\}^2$$

$$= \sum_{ab \in E(G), x \in U} \{2d(x) + (d(a) + d(b) - 2)\}^2$$

$$= 4 \sum_{ab \in E(G), x \in U} d(x)^2 + \sum_{ab \in E(G), x \in U} \{d(a) + d(b) - 2\}^2 + 4 \sum_{ab \in E(G), x \in U} d(x)\{d(a) + d(b)\}$$

$$- 8 \sum_{ab \in E(G), x \in U} d(x).$$

Finally, combining the contributions of $H_1$ and $H_1$, the desired result of theorem 1 follows.

Now we consider reformulated first Zagreb index of a particular generalised hierarchical product graph, where $G = P_2$, as some chemical structures are in this form.

**Corollary.1** The reformulated first Zagreb index of cluster product of two graphs G and $P_2$, with root vertex of degree one is given by

$$EM_1(P_2 \Pi H(U)) = 2EM_1(H) + 8\sum_{u \in U} d(u)^2 - 6\sum_{u \in U} d(u) + 4|\{xy \in E(H) : x, y \in U\}| + 4\sum_{u \in U} \sum_{x \in N[u]} d(x).$$

The Cartesian product of two graphs $G$ and $H$ denoted by $G \times H$ is the graph with the vertex set $V(G) \times V(H)$, with vertices $u = (u_1, u_2)$ and $v = (v_1, v_2)$ connected by an edge if and only if $[u_1 = v_1 \text{ and } (u_2, v_2) \in E(G_2)]$ or $[u_2 = v_2 \text{ and } (u_1, v_1) \in E(G_1)]$. From definition it is clear that the generalised hierarchical product of graph is a subset of Cartesian product of graphs. Also, if $U = V(H)$, then $G \Pi H(U) = G \times H$. Thus from Theorem 1 the following result follows:

**Theorem.2** The reformulated first Zagreb index of Cartesian product of two graphs G and H is given by
$$EM_1(G \times H) = |V(G)| EM_1(H) + |V(H)| EM_1(G) + 12|E(G)| M_1(H)$$
$$+ 12|E(H)| M_1(G) - 32|E(G)||E(H)|.$$

Note that, the exact formula for reformulated first Zagreb index of the Cartesian product of graphs was obtained directly in [10], which coincides with the above result. For different special graphs obtained by specializing components of the Cartesian product of graphs such as $C_4$–nanotube $(P_n \times C_m)$, $C_4$–nanotorus $(C_n \times C_m)$, $n$-prism $(K_2 \times C_n)$, ladder graph $(P_2 \times P_{n+1})$, grid graph $(P_n \times P_m)$ rook's graph $(K_n \times K_m)$ are discussed there in.

## 2.1. The cluster product of graphs

The cluster product of two graphs $G$ and $H$ denoted by $G\{H\}$ is the graph obtained by taking one copy of $G$ and $V(G)$ copies of a rooted graph $H$, and by identifying the root of the $i$-th copy of $H$ with the $i$-th vertex of $G$, for $i = 1, 2, ..., |V(G)|$.

The cluster of two graphs $G$ and $H$ with root vertex $x$, can be represented as a special case of generalised hierarchical product of graphs if $U = \{x\}$, a singleton. Thus we have, $G\Pi H(U) = G\{H\}$. The cluster product of two graphs $G$ and $H$ with root $x$ is also termed as (standard) hierarchical product and is denoted by $G\Pi H (= G\{H\})$. Thus we have, $G\Pi H(U) = G\{H\} = G\Pi H$, if $U = \{x\}$. Thus from Theorem 1, by considering $U = \{x\}$, we get the following result directly.

**Theorem.3** The reformulated first Zagreb index of cluster product of two graphs $G$ and $H$, with root vertex $x$ of $H$ is given by

$$EM_1(G\{H\}) = |V(G)| EM_1(H) + EM_1(G) + 5d(x)M_1(G) + 8|E(G)|d(x)^2$$
$$+ 4|E(G)|\delta(x) - 16|E(G)|d(x).$$

Let $G = (V(G), E(G))$ be a $r$-regular graph, then $d(x) = r$, $\delta(x) = r^2$, $M_1(G) = r^2 |V(G)|$ and $HM(G) = (2r-2)^2 |E(G)| = 2r(r-1)^2 |V(G)|$. Then the following result follows directly from above.

**Corollary.2** If $G$ and $H$ be $r$ and $s$-regular graph, respectively, then

$$EM_1(G\{H\}) = |V(G)| \{2s(s-1)^2 |V(H)| + 2r(r-1)^2 + 5r^2 s + 6rs^2 - 8rs\}.$$

Using the above result, the following example follows:

**Example.1**
(i) $EM_1(C_n\{C_m\}) = 4n(m+15)$
(ii) $EM_1(K_n\{K_m\}) = n[2m(n-1)(n-2)^2 + 2(m-1)(m-2)^2 + 5(n-1)^2(m-1) + 6(n-1)(m-1)^2$
$- 8(n-1)(m-1)]$.

The t-thorny graph or the t-fold bristled graph of a graph $G$ is obtained by joining t number of degree one vertices or thorn to each vertex of $G$. This type of graph was introduced by Gutman [30] and studied by different authors also [31-33]. From construction it is clear that the $t$-thorny graph is the cluster product of $G$ and a star graph $S_{t+1}$, where the root vertex is the central vertex of $S_{t+1}$ with degree $t$. Hence, using theorem 3 we get the following result.

**Corollary.3** Let $G$ be a connected graph with $n$ vertices and $m$ edges, then

$$EM_1(G\{S_{t+1}\}) = EM_1(G) + 5tM_1(G) + nt(t-1)^2 + 8mt^2 - 12mt.$$

We have derived the above result in [10] also as a special case of corona product of graphs and hence different results for some particular thorn graphs were also derived there. Let us now consider one important type of graph which is the cluster product of G and $P_2$. Using theorem 3, its first reformulated Zagreb index is calculated as follows:

**Corollary.4** Let $G$ be a connected graph with $n$ vertices and $m$ edges, then

$$EM_1(G\{P_2\}) = EM_1(G) + 5M_1(G) - 4|E(G)|.$$

## 3. APPLICATION OF GENERALISED HIERARCHICAL PRODUCT AND CLUSTER PRODUCT OF GRAPHS

We know that, many chemically interesting molecular graphs and nanostructures are obtained from different graph operations of some particular graphs, so in this section we investigate such particular molecular graphs which are resultant of generalised hierarchical product and cluster product of some particular graphs. Thus, we apply the results derived in the previous section to compute reformulated first Zagreb index of some special chemically interesting molecular graphs and also of nanostructures. First we consider some chemical structure in the form $P_2 \Pi H(U)$.

**Example.2** The molecular graph of truncated cube can be considered as generalised hierarchical product of $P_2$ and $H(U)$ (shown in figure 1), where $U = \{v_1, v_4, v_9, v_{12}\}$. Since, $EM_1(H) = 200$, thus using Corollary 1, we get $EM_1(P_2 \Pi H(U)) = 576$.

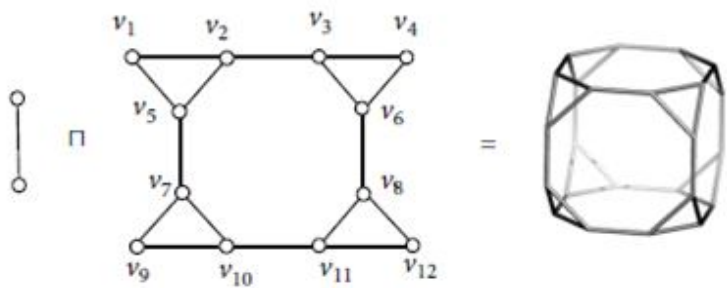

*Figure. 1 The molecular graph of truncated cube*

**Example.3** The dimer fullerene is the generalised hierarchical product of $P_2$ and $C_{60}(U)$, where $U = \{5,6\}$ ( as shown in figure 2). From direct calculation, we have $EM_1(C_{60}) = 1440$. Thus, using Corollary 1, we get $EM_1(P_2 \Pi C_{60}(U)) = 3064$.

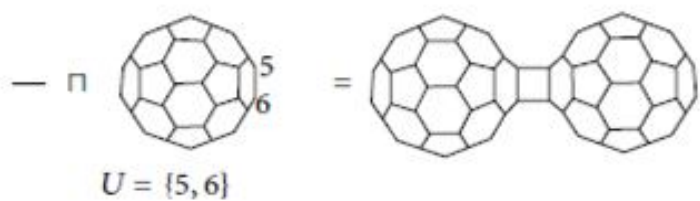

*Figure.2 The dimer fullerene $C_{60}$*

**Example.4** Let $L_n$ be the hexagonal chain with $n$ hexagons (shown in figure 3). From definition it is clear that $L_n = P_2 \Pi P_{2n+1}(U)$, where $U = \{v_1, v_3, v_5, ..., v_{2n+1}\}$. Then using Corollary 3, we get
$$EM_1(L_n) = EM_1(P_2 \Pi P_{2n+1}(U)) = 52n - 28.$$

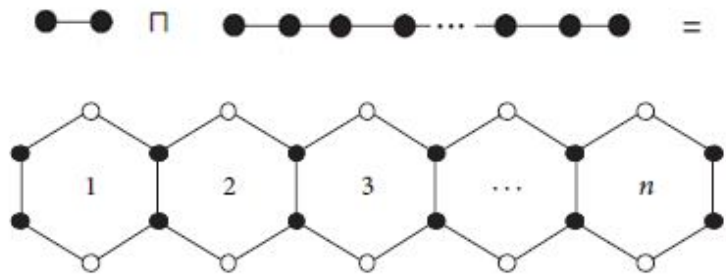

*Figure.3 The hexagonal chain with n hexagons $L_n$*

**Example.5** The zig-zag polyhex nanotube $TUHC_6[2n,2]$ is the generalised hierarchical product of $P_2$ and $C_{2n}(U)$, where $U = \{v_2, v_4, ..., v_{2n}\}$ (figure 4). Thus using Corollary 1, we get
$$EM_1(TUHC_6[2n,2]) = EM_1(P_2 \sqcap C_{2n}(U)) = 52n.$$

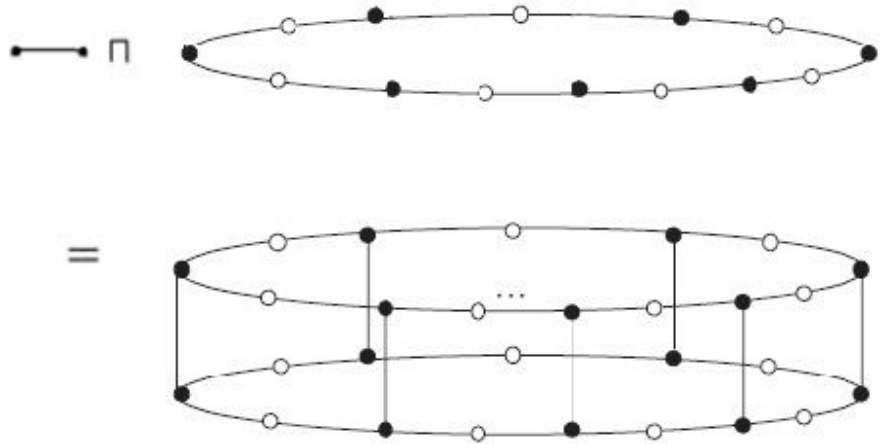

*Figure. 4 The zig-zag polyhex nanotube $TUHC_6[2n,2]$*

**Example.6** The linear phenylene $F_n$ with $n$ benzene ring is the graph given by $P_2 \sqcap P_{3n}(U)$, where $U = \{v_{3k+1} : 0 \le k \le n-1\} \cup \{v_{3k} : 1 \le k \le n\}$. Then using Corollary 1, we get the following result
$$EM_1(F_n) = EM_1(P_2 \sqcap P_{3n}(U)) = 96n - 72.$$

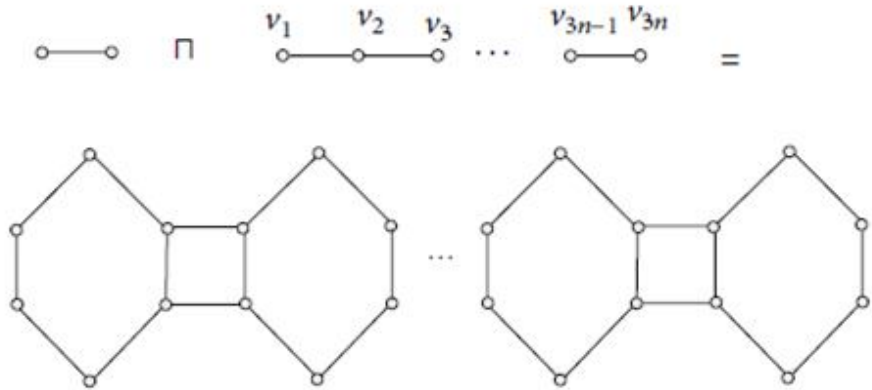

*Figure.5 The linear phenylene $F_n$*

In the following now we consider some important molecular structures which can be obtained as a result of cluster product of some special graphs. First let us consider one important nanostructure called regular dicentric dendrimers.

**Example.7** Let $DD_{p,r}$ denote the regular dicentric dendrimers. Then $DD_{p,r} = P_2\{H\}$, where H is a dendron tree of progressive degree p and generation r (see Figure 6). Thus using theorem 3, we get

$$EM_1(DD_{p,r}) = 2EM_1(H) + 12p^2 - 2p.$$

Since from direct calculation, we have

$$EM_1(H) = \frac{p(p^{r+2} + 3p^{r+1} - 8p^2 + 5p - 1)}{p - 1}.$$

So, we get from above

$$EM_1(DD_{p,q}) = \frac{2p^2(p^{r+1} - 3p^r - 2p - 2)}{p - 1}.$$

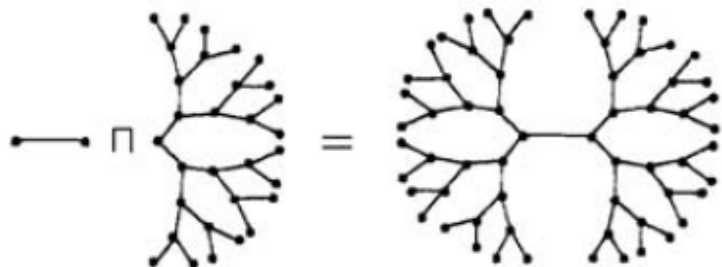

*Figure. 6 The regular dicentric dendrimers $DD_{p,r}$*

**Example.8** The sun graph denoted by $Sun_{(m,n)}$ is defined as cluster product of $C_m$ and $P_{n+1}$, where $P_{n+1}$ is rooted at a vertex of degree one. Then using theorem 3, the reformulated first Zagreb index of $Sun_{(m,n)}$ is given by

$$EM_1(Sun_{(m,n)}) = 2m(2n + 9).$$

**Example:9** Let $\Gamma$ be the molecular graph of octanitrocubane (see Figure 7). It can be considered as the cluster product of a cube (G) and $P_2$. Then using corollary 4, we similarly get $EM_1(\Gamma) = 504$.

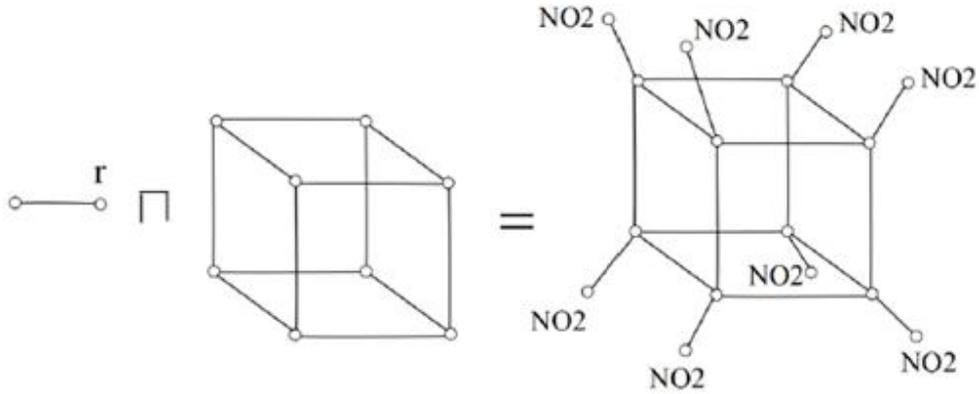

*Figure.7 The Molecular Graph of Octanitrocubane*

**Example.10** The square comb lattice graph $C_q(N)$ with open ends, can be considered as the cluster product $P_n\{P_n\}$, where $N = n^2$ is the total number of vertices of $C_q(N)$ (see Figure 8 ). Here root of $P_n$ is on its pendent vertex i.e. the vertex of degree one. Thus applying theorem 3 we get, for $n \geq 3$

$$EM_1(C_q(N)) = n(8n - 14) + 6(n - 1) + 6(2n - 3) + 6n - 14 = 8n^2 + 10n - 38.$$

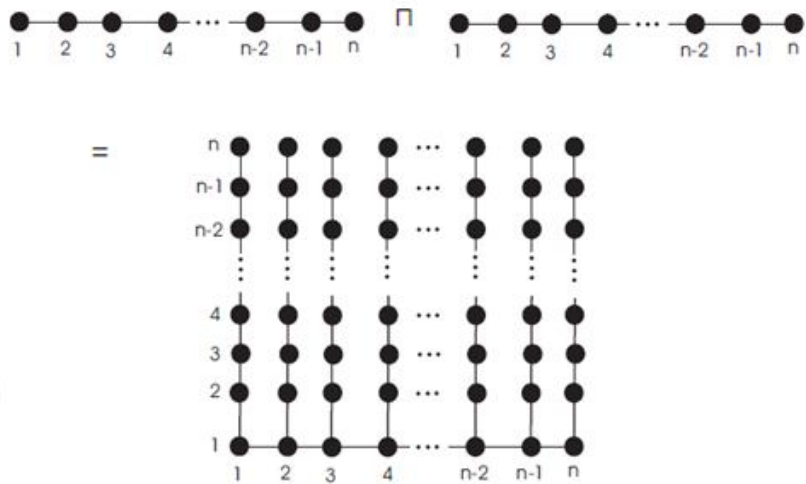

*Figure. 8 The square comb lattice graph $C_q(N)$*

## 4. CONCLUSION

In this paper, we first derive explicit expression of reformulated first Zagreb index of generalized hierarchical product of two connected graphs. Hence using this result we obtain the reformulated first Zagreb index of cluster product of two graphs. Finally, using the derived results the reformulated first Zagreb index of some chemically important graphs such as square comb lattice, hexagonal chain, molecular graph of truncated cube, dimer fullerene etc. are obtained.